# Reformulating the Governing Equations from the Total Energy of a Fluid Parcel: From Internal, Kinetic, and Flow-Work to Stagnation-Enthalpy Balance

Peng SHI[1 *]

[1] Logging Technology Research Institute, China National Logging Corporation, No.50 Zhangba five road, Xi'an, 710077, P. R. China

[*]Corresponding author. Email: sp198911@outlook.com;

**Short title: Stagnation-Enthalpy Balance and Flow Work**

**Abstract**. The classical governing equations of fluid mechanics assign pressure two distinct roles: elastic energy of a fluid parcel in acoustics and a surface force on a control-volume boundary in flow dynamics. This dual role leads to energy equations with different physical content and predictions. This study re-examines the governing equations by applying the first law of thermodynamics directly to a fluid parcel in the Eulerian description, whose energy is taken to comprise internal, kinetic, and flow-work terms. Flow work is reinterpreted as elastic potential energy stored within the parcel rather than as external boundary work. The resulting total-energy equation can be written as a stagnation-enthalpy balance, with parcel energy per unit volume $\rho h_0$, and decomposes into mechanical and thermal parts. The mechanical part is equivalent to the momentum equation, whereas the thermal part is a modified internal-energy equation that retains an explicit pressure material-derivative contribution. In the inviscid adiabatic limit, the steady case recovers the classical stagnation-enthalpy conservation along streamlines, while the unsteady case retains a balance involving $\partial P/\partial t$. The formulation is assessed with the classical Sod shock tube problem by comparing two one-dimensional Euler systems that differ only in their energy-conservation law. The new $h_0$-system is hyperbolic and classical-like, predicts star-region states differing from the traditional results at the few-percent level, and has the total stagnation-enthalpy of tube $\int \rho h_0 dx$ conserved while the traditional total energy grows by2.31%. This conservation duality provides a computable signature of internalized flow work and makes the thermo-mechanical coupling explicit at the equation level.

## 1. Introduction

The study of fluid dynamics is an ancient pursuit: the systematic investigation of fluids can be traced back to Archimedes (287–212 B.C.), whose hydrostatics established the principle of buoyancy and the equilibrium of floating bodies [1, 2]. The modern mathematical description of fluid flow began with Euler, who formulated the equilibrium of fluids in 1752 and, in 1755, expressed the motion of inviscid fluids in terms of partial differential equations [1, 3, 4]. Incorporating Newton's law of internal friction, Navier and Stokes independently extended these equations to account for viscosity, giving what are now known as the Navier–Stokes equations—the fundamental governing equations of fluid dynamics [5, 6]. Today, fluid mechanics underpins nearly every branch of engineering and science concerned with fluid motion, Among its governing equations, the energy equation is of particular importance for heat transfer: it is the vehicle through which viscous dissipation, pressure work, and heat conduction enter the description of a flow, and it thereby governs the coupling between mechanical work and thermal effects that determines convective heat transfer in gas turbines, heat exchangers, electronic cooling, and other energy systems. The continued refinement of the governing equations—and of the energy equation in particular—remains a matter of both fundamental and practical importance.

Fluid mechanics is commonly considered a branch of classical mechanics: the motion of a fluid parcel is treated analogously to that of an ordinary particle and obeys Newton's laws of motion or equivalent mechanical principles [2, 6]. A flow is described in two complementary ways—the Eulerian description, which observes flow quantities at fixed spatial points, and the Lagrangian description, which follows individual fluid parcels through space and time. The Navier–Stokes equations are conventionally derived by applying momentum conservation to a fluid parcel in the Lagrangian framework and are then expressed in field variables, a characteristic feature of the Eulerian description.

Within this framework, the energy aspects of a flow are conventionally treated as follows. The equation of motion is multiplied by the velocity to yield the mechanical-energy equation, and the internal-energy equation—the basis of convective heat-transfer analysis—is obtained either from the first law applied to the fluid or by subtraction from the total-energy balance, the total energy of a fluid element being the sum of its internal and kinetic energy [2, 6, 7]. In the internal-energy equation, the viscous dissipation term appears as the irreversible channel from mechanical energy to heat: its treatment has been a classical concern of convection analysis since the study of natural convection by Gebhart [8], and it continues to receive attention in laminar channel flows [9] and in convection problems where the pressure work acts alongside the viscous dissipation [10]. The irreversible production associated with viscous dissipation is likewise the core quantity of entropy-generation minimization, a thermodynamic design principle for heat-transfer systems [11]. In gas dynamics, this accounting is complemented by the stagnation enthalpy: for inviscid, adiabatic flow, $h_0 = h + \frac{1}{2}U^2$, with $h = e + P/\rho$ the specific enthalpy, is conserved along streamlines, and in unsteady flow the balance generalizes to $\rho Dh_0/Dt = \partial P/\partial t$, in which the local pressure derivative acts as a source term [7, 12, 13].

In acoustics, pressure is regarded as a measure of elastic potential energy, with the potential energy associated with compression represented by the second-order quantity $p^2/(2\rho c^2)$ [14]—an accounting whose engineering extension is thermoacoustics, in which pressure oscillations serve as the working medium that converts heat into mechanical energy and back [15]. In fluid dynamics, by contrast, it is regarded as an external force doing work—that is, as a surface force acting on the boundary of a control volume [16–19]. Although these two perspectives may appear compatible in certain limits, they are fundamentally different in how they place pressure within the energy budget: the former treats it as an intrinsic property of a fluid parcel, representing energy stored through reversible compression, whereas the latter treats it as boundary work exchanged with the surroundings—a boundary interaction or external mechanical action, rather than energy stored within the parcel. Pressure is thus assigned two different roles—as a measure of elastic energy in acoustics and as a surface force on a control-volume boundary in fluid dynamics—and the two roles lead to fluid energy equations with different physical content and different predictions. This contrast motivates the question addressed in this study: what form do the governing equations take if the energy of a fluid parcel is written from the outset as a single thermodynamic quantity comprising internal, kinetic, and flow-work (elastic) terms?

The present study re-examines the governing equations from this viewpoint. Rather than treating flow work as boundary work exchanged with the surroundings, it applies the first law of thermodynamics directly to a fluid parcel in the Eulerian description, with the parcel energy taken to comprise internal, kinetic, and flow-work terms. Flow work is reinterpreted as the elastic potential energy stored within the parcel, so that the total energy per unit volume is identified with the stagnation enthalpy $\rho h_0$. The resulting total-energy equation decomposes into a mechanical part, equivalent to the momentum equation, and a thermal part that differs from the conventional internal-energy equation by retaining an explicit pressure material-derivative contribution. In the inviscid adiabatic limit, the steady case recovers the classical stagnation-enthalpy conservation along streamlines, while the unsteady case retains an explicit balance involving $\partial P/\partial t$. This formulation is not merely an enthalpy-based restatement of the conventional derivation; it makes the mechanical–thermal coupling explicit at the equation level and provides a consistent parcel-energy accounting in which the pressure gradient is interpreted as the spatial manifestation of reversible conversion among internal, flow-work, and kinetic energy, with viscous dissipation providing the irreversible channel. Its physical consequences are examined through a Sod shock tube comparison between the new $h_0$-system and the traditional $E$-system.

The remainder of this paper is organized as follows. Section 2 reviews the material derivative and establishes the notation for expressing Lagrangian rates of change in Eulerian variables. Section 3 derives the kinetic-energy balance of a fluid parcel and shows its equivalence to the momentum conservation equation. Section 4 develops the total-energy conservation equation, identifies the parcel energy with its stagnation enthalpy, decomposes the total-energy equation into mechanical and thermal parts, and derives the corresponding modified internal-energy equation. Section 5 presents the Sod shock tube comparison, including the two model systems, wave structure, thermodynamic states, and the conservation duality of the two energy formulations. The final section summarizes the conclusions of this study.

## 2. Material Derivative

Fluid dynamics is concerned with the state and evolution of the flow field rather than with those of an individual fluid parcel, and it therefore adopts the Eulerian description. The changes experienced by a moving fluid parcel, however, are most naturally stated in the Lagrangian description. The material derivative is the bridge between the two, and it is the vehicle by which the parcel conservation laws in the following sections are written in field variables. We first review its establishment briefly.

When a fluid parcel moves through a time-varying field, the value of the physical quantity it carries at a given position varies with time. Treating the quantity carried by the parcel as a function of time alone, its Lagrangian rate of change can be expressed in the Eulerian description as:

$$\frac{d\boldsymbol{a}(t,\boldsymbol{x}_1)}{dt}=\frac{\boldsymbol{A}(t,\boldsymbol{x})-\boldsymbol{A}(t-\delta t,\boldsymbol{x}-\delta\boldsymbol{x})}{\delta t} \tag{1}$$

here, $\boldsymbol{x}$ is the displacement vector, $\delta\boldsymbol{x}$ is the displacement of a parcel during $\delta t$, $d/dt$ is the time derivative, $\nabla$ is the gradient operator, $\boldsymbol{a}(t, \boldsymbol{x}_1)$ denotes the physical quantity of the parcel that was initially at position $\boldsymbol{x}_1$ and reaches the given point $\boldsymbol{x}$ at time $t$, $\boldsymbol{A}(t, \boldsymbol{x})$ is the physical quantity expressed in Eulerian description varying with time and position.

The quantity described in Lagrangian description that is only a time-dependent function can be seen as a composite function with position as the intermediate variable.

Consequently, for a given parcel, the two forms are related as follows:

$$\boldsymbol{a}(t, \boldsymbol{x}_1) = \boldsymbol{A}(t, \boldsymbol{x}(t)) \tag{2}$$

Thus, Equation (1) is expressed as:

$$\frac{d\boldsymbol{a}(t, \boldsymbol{x}_1)}{dt} = \frac{\partial \boldsymbol{A}(t, \boldsymbol{x})}{\partial t} + \boldsymbol{U}(t, \boldsymbol{x}) \cdot \nabla \boldsymbol{A}(t, \boldsymbol{x}) \tag{3}$$

where $\partial/\partial t$ is the time partial derivative, $\boldsymbol{U}(t, \boldsymbol{x})=\delta \mathbf{x}/\delta t$ is the velocity vector in Eulerian perspective. Accordingly, the material derivative, which expresses the Lagrangian rate of change in Eulerian variables, is written as:

$$\frac{D}{Dt} = \frac{\partial}{\partial t} + \boldsymbol{U} \cdot \nabla \tag{4}$$

where, $D/Dt$ is the material derivative, which expresses the rate of change of a quantity following a fluid parcel in terms of Eulerian field quantities. All parcel conservation laws below are expressed with this operator.

## 3. Kinetic Energy Change of a Fluid Parcel in the Eulerian Description

According to classical mechanics, the kinetic energy of a fluid parcel equals one half of the product of its mass and the square of its velocity. For a given parcel the mass is constant, so its kinetic energy changes only when the parcel accelerates. In the Eulerian description, the rate of change experienced by a specific fluid parcel is expressed by the material derivative introduced above.

The rate of change of the kinetic energy of a parcel equals the power of the net force acting on it, namely the product of the net force and the velocity of the parcel. Expressed in the Eulerian description, the kinetic-energy balance of a fluid parcel reads

as:

$$\frac{1}{2}\frac{D\left(\rho VU^{2}\right)}{Dt}=\boldsymbol{F}\cdot\boldsymbol{U} \tag{5}$$

where $\rho$ is the density of fluid parcel, $V$ is the volume of a fluid parcel passing through the point, $U$ is the module of $\boldsymbol{U}$, $\boldsymbol{F}$ is the resultant force in Eulerian perspective.

Applying the product rule to the material derivative, the term on the left side of Equation (5) can be rewritten as:

$$\frac{1}{2}\frac{D\left(\rho VU^{2}\right)}{Dt}=\frac{1}{2}\frac{D\left(\rho V\right)}{Dt}U^{2}+\frac{1}{2}\rho V\frac{D\left(U^{2}\right)}{Dt} \tag{6}$$

For a specific fluid parcel, the density and the volume may change during the motion, but the mass does not vary with time or spatial position, so the following equation holds true [2]:

$$\frac{D\left(\rho V\right)}{Dt}=\rho\frac{DV}{Dt}+V\frac{D\rho}{Dt}=0 \tag{7}$$

Then, the rate of change of fluid kinetic energy in the Eulerian description can be stated as:

$$\frac{1}{2}\frac{D\left(\rho VU^{2}\right)}{Dt}=\frac{1}{2}\rho V\left(\frac{\partial\left(U^{2}\right)}{\partial t}+\boldsymbol{U}\cdot\nabla\left(U^{2}\right)\right) \tag{8}$$

Since the following identities hold:

$$\boldsymbol{U}\times\left(\nabla\times\boldsymbol{U}\right)\cdot\boldsymbol{U}=0 \tag{9}$$

$$\boldsymbol{U}\cdot\nabla\boldsymbol{U}=\frac{1}{2}\nabla\left(U^{2}\right)+\left(\nabla\times\boldsymbol{U}\right)\times\boldsymbol{U} \tag{10}$$

With Equations (8), (9) and (10), Equation (5) can be rewritten as:

$$\rho V\left(\frac{\partial\boldsymbol{U}}{\partial t}+\boldsymbol{U}\cdot\nabla\boldsymbol{U}\right)\cdot\boldsymbol{U}=\boldsymbol{F}\cdot\boldsymbol{U} \tag{11}$$

From Equation (4), the term in parentheses on the left-hand side of Equation (11) corresponds to the acceleration of a fluid parcel in the Eulerian description. That is, the energy conservation described by Equation (11) is equivalent to the momentum conservation.

Assuming that the external forces acting on the fluid do conservative work, the work done by the external forces can be expressed using a scalar potential function as:

$$\frac{1}{2}\rho V\frac{D\left(U^2\right)}{Dt}=\frac{D\Phi}{Dt} \tag{12}$$

where $\Phi$ represents the potential energy, which is generalized to the unsteady case here. The terms on the right-hand side of Equation (12) represent the power of the potential energy. The mechanical energy conservation described by Equation (12) differs from that in classical mechanics. This difference arises because classical mechanics generally treats conservative force fields as steady fields, whereas flow fields are often unsteady. This further suggests that classical mechanics has an inadequate understanding of potential energy and the work done by it.

Traditionally, the equations of motion of a fluid are obtained by substituting the forces acting on the fluid into the momentum conservation equation. During fluid flow, however, energy conversion occurs constantly: kinetic energy is converted into internal energy by viscous dissipation and exchanges with flow work through pressure. Although momentum conservation is a fundamental law of classical mechanics, a derivation based solely on momentum conservation does not make this conversion explicit. Equation (12) shows that the momentum equation is equivalent to the kinetic-energy balance expressed by Equation (12). Therefore, the equations of motion can equally be derived from energy conservation, with the advantage that the energy conversion accompanying the flow is made explicit. This provides the basis for the derivation developed in the following section.

## 4. Energy Conservation of Fluid

According to the first law of thermodynamics, the change in the energy of a system equals the heat added to the system plus the work done on the system. The total-energy balance of a fluid parcel written in the Eulerian description, reads as:

$$\frac{DE}{Dt}=\frac{DQ}{Dt}+\frac{DW}{Dt} \tag{13}$$

where, $E$ is the total energy, $Q$ and $W$ are the heat added to and the work done on the parcel by its surroundings, respectively. For a fluid parcel, $E$ is taken to be the sum of the internal energy, the flow-work (elastic potential energy), and the kinetic energy of the parcel.

The rate of change of the total energy of a fluid parcel is expressed in the Eulerian description as:

$$\frac{DE}{Dt}=\rho V\frac{De}{Dt}+\frac{D(PV)}{Dt}+\frac{1}{2}\rho V\frac{D(U^2)}{Dt} \tag{14}$$

where $e$ is the internal energy per unit mass, a field quantity depending on time and the spatial coordinates. The three terms on the right-hand side of Equation (14) correspond, in order, to the rates of change of internal energy, flow-work, and kinetic energy of the fluid under the Eulerian description. Since the elastic potential energy per unit volume is the pressure itself—precisely the flow-work term of thermodynamics—the sum of the three energy forms can be written compactly as $\rho(e + P/\rho + ½U^2) = \rho h_0$, where $h =$ $e + P/\rho$ is the specific enthalpy and $h_0 = h + ½U^2$ is the stagnation enthalpy. The total energy carried by a fluid parcel is thus identified with its stagnation enthalpy. With Equation (14), the total energy of a fluid parcel of unit volume $\hat{E}$ is expressed as:

$$\frac{D\hat{E}}{Dt} = \rho\frac{De}{Dt} + \rho\frac{D}{Dt}\left(\frac{P}{\rho}\right) + \rho\frac{D}{Dt}\left(\frac{U^2}{2}\right) \tag{15}$$

Splitting the flow work in Equation (15) into two terms yields:

$$\frac{D\hat{E}}{Dt} = \rho\left(\frac{De}{Dt} + P\frac{D}{Dt}\left(\frac{1}{\rho}\right)\right) + \left(\rho\frac{D}{Dt}\left(\frac{U^2}{2}\right) + \frac{DP}{Dt}\right) \tag{16}$$

For an adiabatic process of a fluid parcel with no external work, Equation (16) represents the energy conservation equation for the fluid parcel. The first and second terms on the right-hand side of Equation (16) correspond to the conservation of internal energy and mechanical energy, respectively. In this process, the elastic potential energy (flow work) is resolved into two contributions: the work done by external forces, which changes the internal energy of the fluid parcel, and the work done by potential forces, which changes its kinetic energy.

When a fluid flows, its fluid parcels contact with one another, interactions occur among the fluid parcels, and energy is transferred and converted between them. Considering the interactions between fluid parcels and substituting the conventional heat transfer and external work into Equation (16), the total energy conservation equation for a fluid parcel is expressed as follows:

$$\rho\frac{De}{Dt} + \rho\frac{D}{Dt}\left(\frac{P}{\rho}\right) + \rho\frac{D}{Dt}\left(\frac{U^2}{2}\right) = \nabla\cdot(k\nabla T) + \nabla\cdot(\boldsymbol{\sigma}\cdot\boldsymbol{U}) - \nabla\cdot(P\boldsymbol{U}) \tag{17}$$

where $\boldsymbol{\sigma}$ is the viscous stress tensor, a symmetric second-order tensor, $k$ is the coefficient of thermal conductivity, $T$ is the temperature. The terms on the right-hand side of Equation (17) represent the heat absorbed by the fluid parcel and the work done on it by external forces. Meanwhile, the fluid parcel satisfies the classical momentum conservation:

$$\rho\frac{\partial \boldsymbol{U}}{\partial t}+\rho\boldsymbol{U}\cdot\nabla\boldsymbol{U}+\nabla P-\nabla\cdot\boldsymbol{\sigma}=0 \tag{18}$$

With Equations (11) and (18), the internal energy conservation equation can be obtained from Equation (17) and be expressed as:

$$\rho\frac{De}{Dt}+\rho\frac{D}{Dt}\left(\frac{P}{\rho}\right)=\nabla\cdot\left(k\nabla T\right)+\boldsymbol{\sigma}:\nabla\boldsymbol{U}-P\nabla\cdot\boldsymbol{U} \tag{19}$$

where : is the double dot product.

Since the following identity holds [2]:

$$\frac{DV}{Dt}=V\nabla\cdot\boldsymbol{U} \tag{20}$$

Substituting Equation (20) into Equation (7), the mass conservation equation is obtained as:

$$\frac{\mathrm{D}\rho}{Dt}+\rho\nabla\cdot\boldsymbol{U}=0 \tag{21}$$

Then, Equation (19) can be rewritten as:

$$\rho\left(\frac{\partial e}{\partial t}+\boldsymbol{U}\cdot\nabla e\right)+\frac{\partial P}{\partial t}+\boldsymbol{U}\cdot\nabla P+2P\nabla\cdot\boldsymbol{U}=\nabla\cdot\left(k\nabla T\right)+\boldsymbol{\sigma}:\nabla\boldsymbol{U} \tag{22}$$

Equation (22) is the new internal-energy equation of the fluid, which differs from the conventional one. In the traditional framework, the total energy of a fluid parcel is taken as the sum of internal and kinetic energy, and flow work enters only as boundary work performed by the surroundings; consequently, an adiabatic closed system on which no external work acts conserves “kinetic energy + internal energy,” and the pressure contribution is not counted as energy stored within the parcel. In the present framework, by contrast, flow work is counted as elastic potential energy stored within the parcel, so that the parcel energy becomes $\rho h_0$ and the corresponding conserved quantity for an adiabatic closed system is the stagnation enthalpy.

In the traditional theoretical framework, flow work is interpreted as external work. Although this framework can describe the flow characteristics of fluids, it cannot explain wave phenomena in fluids as the traditional understanding that wave phenomena is the conversion and propagation between kinetic energy and elastic potential energy. The new theoretical framework avoids this shortcoming. The study treats flow work as the elastic potential energy of the fluid rather than as external work. Within this framework, when the fluid's mechanical energy conservation equation is used to explain wave propagation in fluids, wave phenomena can be reasonably interpreted as the conversion and transmission between the fluid's elastic energy and kinetic energy.

## 5. Sod Shock Tube Simulation Comparison

To assess the physical consequences of the new energy framework developed in Section 4 — in particular the conclusion that, once flow work is treated as elastic potential energy stored inside the fluid parcel, the transported conserved energy of the fluid is its stagnation enthalpy $\rho h_0 = \rho(e + U^2/2 + P)$ rather than the conventional one— the classical Sod shock tube problem is computed with two one-dimensional Euler systems that differ only in their energy-conservation law. All other ingredients (mass and momentum equations, equation of state, numerical method, initial and boundary data) are identical in the two computations, so that every difference in the results is attributable solely to the treatment of flow work. The new model is denoted the $h_0$ system and the traditional one below.

## 5.1 The two model systems and the numerical method

In conservative form the traditional (old) model reads

$$\Pi = \left(\rho,\quad \rho \boldsymbol{U},\quad E_{old}\right)^T,\quad \left(\rho \boldsymbol{U},\quad \rho U^2 + P,\quad U\left(E_{old} + P\right)\right)^T \tag{23}$$

while the new model, consistent with Equations (14)–(22), replaces the conserved energy density $\rho E_{old}$ by $\rho h_0$ and carries the corresponding flux:

$$\Pi = \left(\rho,\quad \rho \boldsymbol{U},\quad \hat{E}\right)^T,\quad \left(\rho \boldsymbol{U},\quad \rho U^2 + P,\quad U\left(\hat{E} + P\right)\right)^T \tag{24}$$

With

$$E = \rho\left(c_v T + \frac{U^2}{2}\right),\qquad \hat{E} = \rho\left(c_v T + \frac{U^2}{2} + \frac{P}{\rho}\right),\qquad P = \rho R T \tag{25}$$

where $c_v$ is the isochoric specific heat capacity. The two systems differ only in the third component. Their quantitative wave structure is, however, different. Linearization of Equation (24) yields a linearly degenerate contact field $\lambda_0 = U$ (across which $U$ and $P$ are continuous) and two acoustic fields $\lambda = U \pm c$ with

$$c^2 = \frac{(2\gamma - 1)P}{\gamma \rho},\qquad \lambda_0 = U,\qquad \lambda^{\pm} = U \pm c \tag{26}$$

where $c$ is the local speed of sound of the new model, $\lambda_0$ is the characteristic velocity of the contact discontinuity, $\lambda^{\pm}$ is the characteristic velocities of the right- and left-running pressure waves, $\gamma$ is the adiabatic index. for $\gamma = 1.4$ the new-model sound speed is about 0.958 times the classical value $\sqrt{(\gamma p/\rho)}$. Combining Equation (24) with the mass and momentum equations gives the pressure balance along a particle,

$$\frac{DP}{Dt} = -\frac{(2\gamma - 1)P}{\gamma} \cdot \frac{\partial U}{\partial x} \tag{27}$$

whose coefficient $(2\gamma - 1)/\gamma$ is smaller than the classical value $\gamma$: compression heating and rarefaction cooling are both weaker in the new model. The isentropic-like path of the rarefaction fan becomes $P \propto \rho(2\gamma-1)/\gamma$, and the Riemann invariant carried by the left-running characteristics of the fan is $U + 2\gamma c/(\gamma - 1)$. Both systems are hyperbolic and possess the same three-wave topology (left rarefaction fan, contact discontinuity, right shock); the new model is therefore classical-like and well posed.

Both systems are solved with an identical finite-volume scheme: MUSCL-minmod reconstruction of the primitive variables, Rusanov fluxes, SSP-RK2 time stepping, CFL number 0.4, 800 cells on the 20 m tube ($x = 0$ at the diaphragm), and reflecting end walls (a closed tube). The initial states are $P_L = 1\times10^5$ Pa, $\rho_L = 1$ kg/m³ and $P_R = 1\times10^4$ Pa, $\rho_R = 0.125$ kg/m³ with $U = 0$, and profiles are output at $t = 12$ ms. With the traditional system the computation reproduces the exact Riemann solution — $P^* = 30.31$ kPa, U* = 293.3 m/s, $\rho_L^* = 0.4263$ kg/m³, $\rho_R^* = 0.2656$ kg/m³ — to within about 0.1%. With the new system it matches the analytic two-wave solution constructed from Equations (26) and (27) — $P^* = 30.83$ kPa and $U^* = 307.7$ m/s — to the same accuracy, which validates the characteristic analysis above.

## 5.2 Wave structure and thermodynamic states

Figure 1 shows the $x$–$t$ maps of the new model together with the wave fronts measured at every time step. The classical three-wave pattern is recovered, and the measured front speeds agree with the analysis (fan head at $-c$, contact at $U^*$, shock at the speed given by the jump conditions). Figure 2 compares the profiles at $t$ = 12 ms, and Table 1 lists the star-region states at $t$ = 12 ms together with the analytic references.

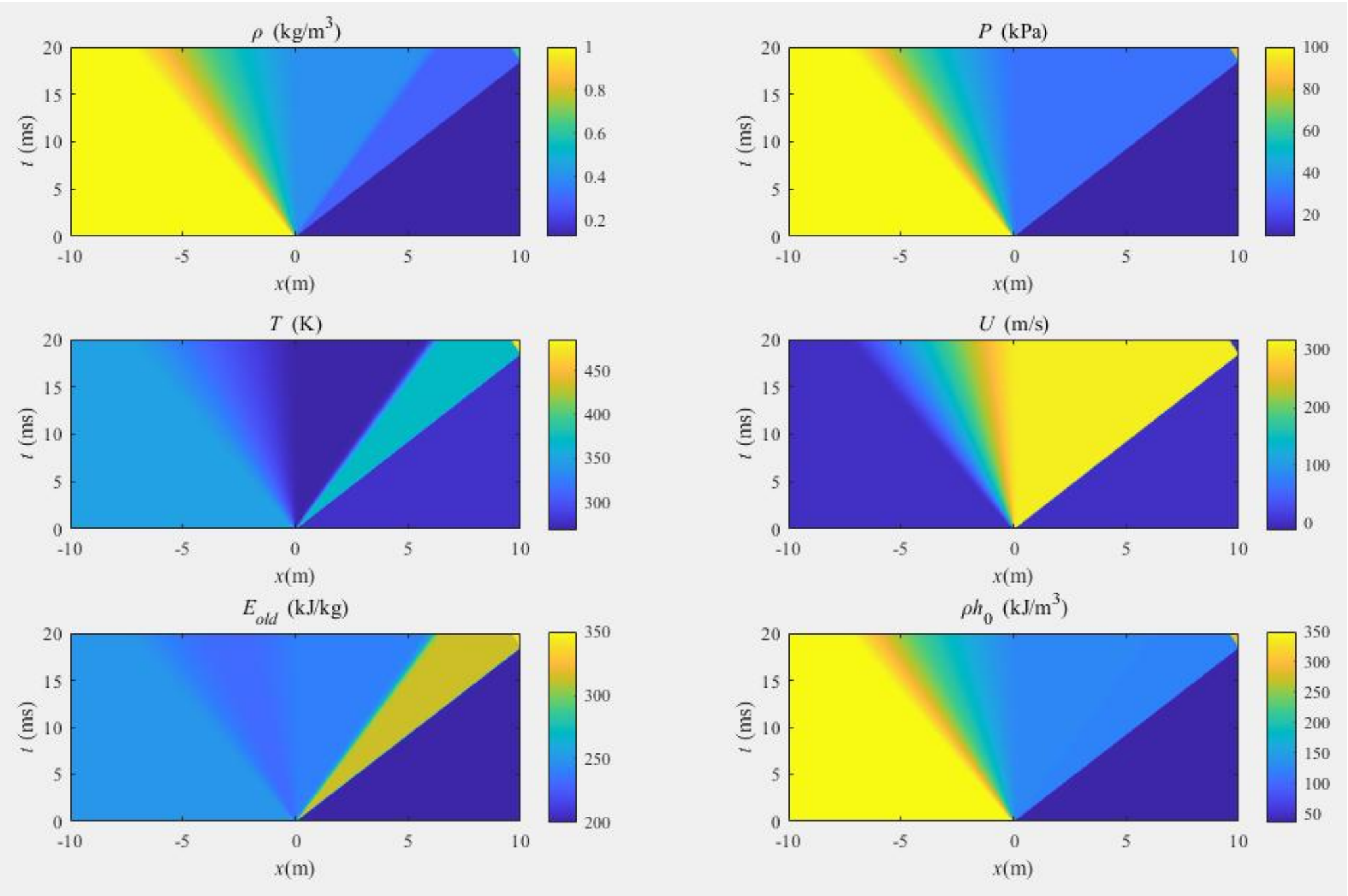

**Figure 1. *x–t* maps of the new model: density, pressure, temperature, velocity, traditional total energy and stagnation-enthalpy ($\rho h_0$).**

**Table 1. Star-region states at *t* = 12 ms**

| Quantity | Classical model | New model |
|---|---|---|
| $P^*$ (kPa) | 30.31 | 30.83 |
| $U^*$ (m/s) | 293.3 | 307.6 |
| $\rho_L^*$ (kg/m³) | 0.4263 | 0.4004 |
| $\rho_R^*$ (kg/m³) | 0.2656 | 0.2894 |
| $T_L^*$ (K) | 247.8 | 268.3 |
| $T_R^*$ (K) | 397.7 | 371.1 |
| Shock position (m) | 6.66 | 6.51 |
| Rarefaction-fan head (m) | −4.66 | −4.46 |

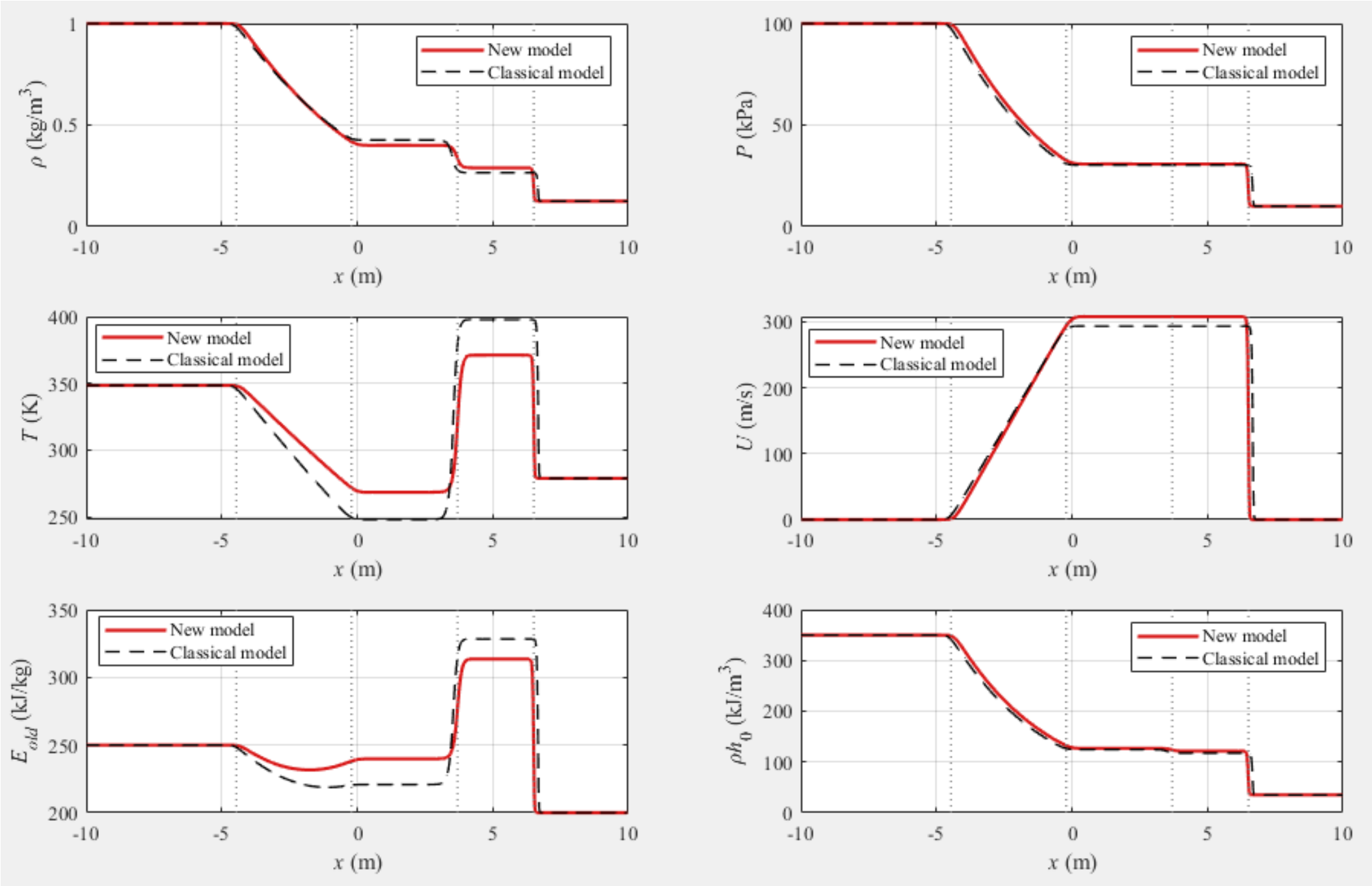

**Figure 2. Flow profiles at t = 12 ms: new model, solid; traditional model, dashed. Vertical dotted lines mark the measured front positions of the new model.**

Compared with the traditional model, the new model gives a slightly higher star pressure (30.83 vs 30.31 kPa, +1.7%) and a higher particle velocity (307.7 vs 293.3 m/s, +4.9%), a slower shock (6.50 vs 6.65 m at 12 ms, consistent with the reduced sound speed in Equation (26)), weaker rarefaction cooling ($T_L$* = 268.5 vs 247.8 K) and weaker shock heating ($T_R$* = 371.2 vs 397.7 K). The shock still raises the temperature of the initially 278.7 K gas, but by less than in the traditional framework — in direct accordance with the 9/7 coefficient in Equation (27) versus $\gamma$.

In a closed tube neither model exchanges mass or work through the walls, so each model conserves exactly the total quantity it postulates as energy — and only that one. Table 2 lists the drift of the tube totals over 0–20 ms and Figure 3 shows the histories. The traditional model has the conventional total energy $\int E_{old}dx$ conserved while $\int \hat{E}dx$ drifts by −2.18%; the new model has $\int \hat{E}dx$ conserved while the conventional total energy $\int E_{old}dx$ increases systematically by +2.31%. This conservation duality is the computable expression of the theoretical difference discussed in Section 4: once flow work is treated as internal elastic energy of the parcel, the transported conserved energy is the stagnation enthalpy $\rho h_0$, and "internal + kinetic" energy alone is not a conserved quantity.

**Table 2. Drift of the tube totals over 0–20 ms.**

| Total quantity | classical model: initial/drift | New model: initial/drift |
|---|---|---|
| Mass (kg) | 11.2500 | 11.2500 |
| Stagnation-enthalpy (J) | $3.850\times10^6$/−2.18% | $3.850\times10^6$ |
| Traditional total energy (J) | $2.750\times10^6$/0 | $2.750\times10^6$/+2.31% |

The comparison therefore demonstrates that (i) the new model is hyperbolic and classical-like, (ii) its quantitative predictions differ from the traditional ones at the few-percent level for this flow, and (iii) the two frameworks are unambiguously distinguishable through which total quantity is conserved. The 2.31% growth of traditional total energy of sod shock tube in the new model, with stagnation-enthalpy ($\rho h_0$) constant, is the measurable signature of the internalized flow work.

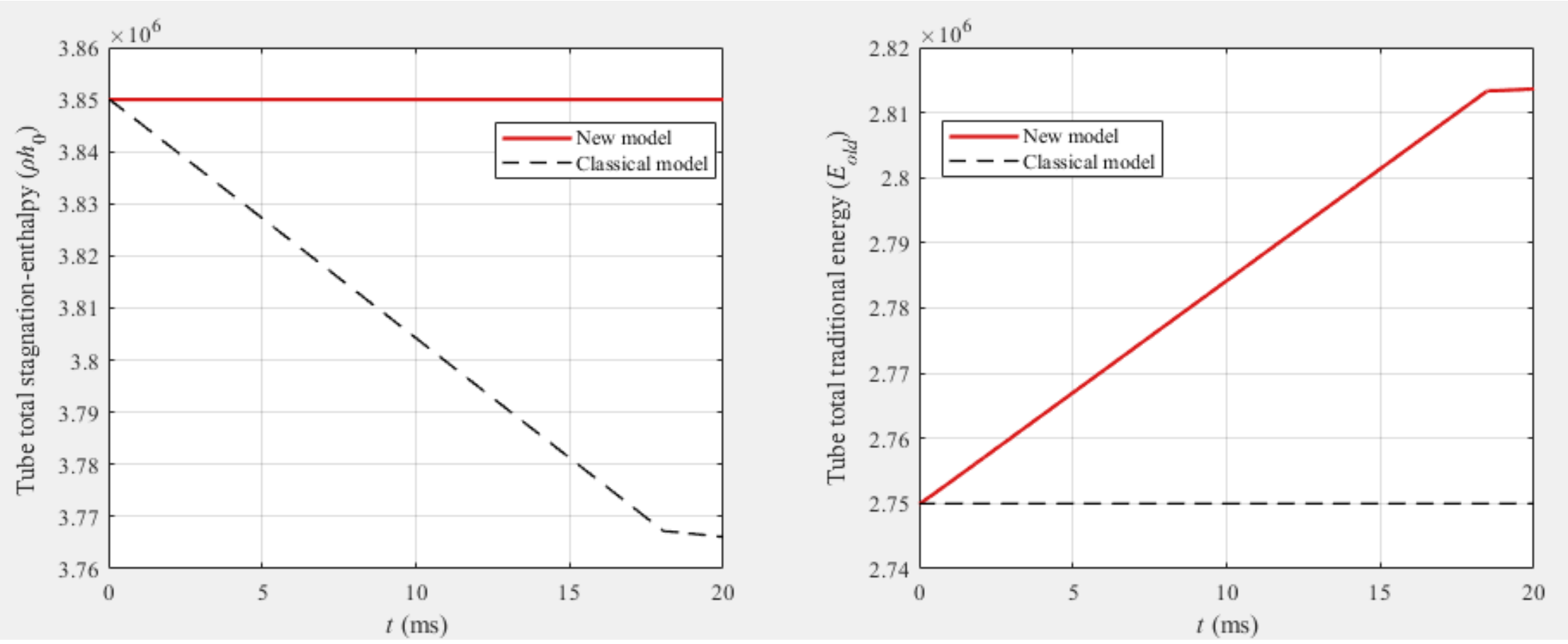

**Figure 3. Tube totals versus time for the new (solid) and traditional (dashed) models: total stagnation-enthalpy ($\rho h_0$) and total traditional energy ($E_{old}$).**

## Conclusion

In this study, the fundamental governing equations of fluid mechanics have been re-derived from the perspective of energy conservation. By defining the total energy of a fluid parcel as the sum of its internal energy, kinetic energy, and flow work, and by reinterpreting flow work as the elastic potential energy stored within the parcel rather than as external boundary work, a total-energy conservation equation has been established in the Eulerian description. This formulation identifies the total energy of a fluid parcel with its stagnation enthalpy ($\rho h_0$) and decomposes the total-energy balance into a mechanical part, equivalent to the momentum equation, and a thermal part that retains an explicit pressure material-derivative contribution relative to the conventional internal-energy equation. The mechanical–thermal coupling is thus made explicit at the equation level. The formulation was assessed with the classical Sod shock tube problem by comparing two one-dimensional Euler systems that differ only in their energy-conservation law. The new $h_0$-system is hyperbolic and classical-like, recovers the classical three-wave topology, and predicts star-region states that differ from the traditional results at the few-percent level. More importantly, the two frameworks are unambiguously distinguishable through the total quantity they conserve: The new model system conserves total stagnation enthalpy, whereas the classical model system conserves total energy (kinetic energy plus internal energy). The present formulation should be regarded as an alternative parcel-energy accounting that clarifies the role of flow work and pressure in unsteady and thermo-mechanical flows, rather than as a replacement of momentum conservation. Its implications for viscous unsteady flows, wave propagation, and convection heat transfer warrant further study.

## Acknowledgments

This work was supported by Deep Earth Probe and Mineral Resources Exploration–National Science and Technology Major Project (2024ZD1003000) and the scientific research and technology development projects of China National Petroleum Corporation (2025DJ104).

## Statements and Declarations

The authors declare that they have no known competing financial interests or personal relationships that could have appeared to influence the work reported in this paper.

## References

[1] J.D. Anderson, A History of Aerodynamics and Its Impact on Flying Machines, Cambridge University Press, Cambridge, 1997.

[2] F.M. White, Viscous Fluid Flow, third ed., McGraw-Hill, New York, 2006.

[3] W. Sheng, A revisit of Navier–Stokes equation, Eur. J. Mech. B Fluids 80 (2020) 60–71.

[4] O. Darrigol, Worlds of Flow: A History of Hydrodynamics from the Bernoullis to Prandtl, Oxford University Press, Oxford, 2005.

[5] H. Schlichting, K. Gersten, Boundary-Layer Theory, ninth ed., Springer, Berlin, 2017.

[6] G.K. Batchelor, An Introduction to Fluid Dynamics, Cambridge University Press, Cambridge, 2000.

[7] H.W. Liepmann, A. Roshko, Elements of Gasdynamics, Wiley, New York, 1957.

[8] B. Gebhart, Effects of viscous dissipation in natural convection, J. Fluid Mech. 14 (1962) 225–232.

[9] A. Barletta, Laminar mixed convection with viscous dissipation in a vertical channel, Int. J. Heat Mass Transfer 41 (1998) 3501–3513.

[10] A. Barletta, D. A. Nield, Effect of pressure work and viscous dissipation in the analysis of the Rayleigh–Bénard problem. Int. J. Heat Mass Transfer, 52 (2009) 3279-3289.

[11] A. Bejan, Entropy Generation Minimization: The Method of Thermodynamic Optimization of Finite-Size Systems and Finite-Time Processes, CRC Press, Boca Raton, 2013.

[12] C.B. Laney, Computational Gasdynamics, Cambridge University Press, Cambridge, 1998.

[13] A.H. Shapiro, The Dynamics and Thermodynamics of Compressible Fluid Flow, The Ronald Press Company, New York, 1953.

[14] A.D. Pierce, Acoustics: An Introduction to Its Physical Principles and Applications, Acoustical Society of America, Woodbury, NY, 1989.

[15] G.W. Swift, Thermoacoustics: A Unifying Perspective for Some Engines and Refrigerators, second ed., Springer, Cham, 2017.

[16] L.D. Landau, E.M. Lifshitz, Fluid Mechanics, second ed., Pergamon Press, Oxford,

1987.

[17] C. Hirsch, Numerical Computation of Internal and External Flows: The Fundamentals of Computational Fluid Dynamics, second ed., Elsevier, Oxford, 2007.

[18] A. Arakawa, Computational design for long-term numerical integration of the equations of fluid motion: two-dimensional incompressible flow. Part I, J. Comput. Phys. 1 (1966) 119–143.

[19] J.B. Perot, Discrete conservation properties of unstructured mesh schemes, Annu. Rev. Fluid Mech. 43 (2011) 299–318.